# Unsupervised paradigm for information extraction from transcripts using BERT

Aravind Chandramouli, Siddharth Shukla, Neeti Nair, Shiven Purohit, Shubham Pandey and Murali Mohana Krishna Dandu

Tredence Analytics Sol. Pvt. Ltd.
No.180, 2nd Floor, Mfar Silverline Tech Park EPIP Zone, Industrial Area, Whitefield, Bengaluru, Karnataka 560066

**Abstract.** Audio call transcripts are one of the valuable sources of information for multiple downstream use cases such as understanding the voice of the customer and analysing agent performance. However, these transcripts are noisy in nature and in an industry setting, getting tagged ground truth data is a challenge. In this paper, we present a solution implemented in the industry using BERT Language Models as part of our pipeline to extract key topics and multiple open intents discussed in the call. Another problem statement we looked at was the automatic tagging of transcripts into predefined categories, which traditionally is solved using supervised approach. To overcome the lack of tagged data, all our proposed approaches use unsupervised methods to solve the outlined problems. We evaluate the results by quantitatively comparing the automatically extracted topics, intents and tagged categories with human tagged ground truth and by qualitatively measuring the valuable concepts and intents that are not present in the ground truth. We achieved near human accuracy in extraction of these topics and intents using our novel approach.

**Keywords.** Key topic detection, Topic modelling, Intent discovery, Conversational noise removal, Unsupervised multi-label tagging, Information retrieval, Natural Language Processing, Unsupervised information extraction, BERT language model

## 1    Introduction

### 1.1    Problem statement

In this paper, we describe an NLP system developed using advanced AI algorithms to extract meta-data from voice transcripts for one of our clients. The voice transcripts here are text converted from the audio recording of conversations between Analysts and corporate customers. The meta-data that our client wanted to extract were key topics, customer intents and classifying these transcripts into pre-defined categories. The meta-data extraction was previously done manually by human experts which was tedious, inefficient, inconsistent, and non-scalable.

Most research and solutions typically approach the above problems using supervised methods [1,2]. However, in an industrial setting, getting tagged data is expensive and usually not available. To overcome these limitations, we introduce TINCE (Topics INtent and Class Extractor), a system that uses an unsupervised approach to help systematically understand the business calls by using its three main components: Key Concepts Identification, Intent Segment Extraction, and Multi-label Document Tagging.

**1.2  Challenges and Objectives**

Processing voice transcripts are typically more challenging than a clean text because they have incorrect sentence segmentation and incorrect words/spellings. Moreover, the lack of speaker text segregation hinders the application of NLP algorithms only to client spoken segment. There are challenges regarding the way people speak during calls such as off-topic conversations that impacts the effectiveness of the algorithms. In this paper, we will discuss the developed conversational casual dialogues identification method that helps in removing the irrelevant off-topic dialogues exchanged between the speakers.

Key concepts are defined as phrases that express the main content of a document. A user can understand the theme of the document by going through the key phrases rather than the entire text document. This process in turn saves a considerable amount of user's time. Consequently, automatic key concept extraction has been widely used to empower many downstream applications. Key concept extraction can be considered as a ranking problem solved by either supervised or unsupervised methods. Since supervised learning requires a large amount of expensive training data, we propose an unsupervised learning key phrase extraction approach in this paper that is scalable and customizable to accommodate any user requirement.

User Intent refers to a sentence or group of sentences that describes the customer intent for calling in. For example, "I want to cancel a reservation", is an example of client Intent. It can help summarize the user objectives and functions present in each call transcript. It can highlight and help prioritize common bugs and issues reported to technical customer support and help in better content creation and planning. We propose an Open Intent Segment Extraction method that automatically extracts user intents in conversational data without pre-defined list of intent classes that the text may contain. It can recognize newly emerging intent segments that it has never seen before. The intent segments solely convey a whole idea along with the necessary context. Tackling such an open world case is much more challenging than the closed-world classification setting predominantly found in literature [3,4,5].

Content-based document labelling or categorization into one or more classes known as Multi-Label Document Classification is extremely useful for organizing available text documents into pre-defined categories [6,7]. This paper will describe an unsupervised approach to tag the client call transcripts into multiple categories by leveraging the text description documents available corresponding to each category.

To summarize, the key contributions of our work are:

- Propose a customizable and scalable unsupervised key phrase extraction method along with explainability feature for the extracted themes.
- Formulate a novel problem of Open Intent Segments Recognition in text. The proposed solution dynamically outputs multiple intents for each transcript in an unsupervised fashion.
- Propose an unsupervised multi-label document tagging method.

## 2  Related work

In this section we focus mainly on the related techniques which are commonly used to solve these kinds of problem. One of the most prominent approaches is to use topic modelling to extract the relevant topics in the textual data. Some of the standard toolkits used for topic modelling are mentioned below:

Another approach to extract the relevant information from the textual data is to use Key phrase extraction (KPE) algorithms which uses combination of rules, heuristics, word level statistics and graph-based algorithms. SGRank [8] is one such algorithm which enables documents to be mapped to a concise set of phrases that can be used for indexing, clustering, ontology building, auto tagging and other information organization schemes. The genesis of algorithms in this space started with TextRank [9] which was a graph based ranking model for text processing where the objective was unsupervised methods for keyword and sentence extraction. The above stated methods mainly focus on getting the key concepts in a domain independent fashion. In the industry setting, clients have very specific definitions of what constitutes a concept and hence, these methods do not provide the desired results. Also, they do not present the explanation behind the extraction of topics making them unsuitable in a business setting.

Previous works have investigated the classification of text into predefined intents. Zhong et al [1] have proposed a convolutional neural network (CNN)-based supervised learning model with word embeddings to extract semantic features in the transcripts for classifying each call into one of the four intent categories. Since there can be multiple intents associated with each transcript, Vedula et al. [10] have introduced the problem of open intent extraction from natural language interaction data. The work also proposes an unsupervised approach using intent indicators to recognize intent in the text. However, the work calls out the challenges to construct a comprehensive list of all intent indictors. We propose generalized intent sentence indicators to identify potential intent sentences that can be combined to form Intent Segments. Moreover, Boakye et al. [11], Shriberg et al. [12] and Stolcke et al. [13] showed that prosodic features were useful for question detection in English conversational speech, but (at least in the absence of recognition errors) most of the performance was achieved with words alone. These are not traditionally used for intent recognition.

Tagging of documents is usually modelled as a supervised multi-class/multi-label classification problem, Mukalov et al [2] in his work shows the use of Neural Networks for article tagging cycle, ranging from data acquisition to tag storing. A common approach to solve the multi-label classification problem is the problem transformation, specifically the binary relevance method [14, 15, 16] in which the multi-label problem will be split into binary classification subtasks. Similar to the limitations of the above problem statements, unavailability of tagged data constrains to use an unsupervised approach to classify the documents. Work done by Łopuszyński et al. [17] tries to solve the problem using an unsupervised approach by extracting noun phrases from the tag description and matching the root form with the text phrases obtained from the corpus which is to be tagged. This approach has a limitation of not capturing other POS patterns and discriminatory tokens which could play a major role in correctly tagging the transcripts.

## 3      Method

### 3.1    Conversational Casual Dialogues Identification

Call transcripts typically have a lot of casual talk such as salutations concentrated in the beginning of the call and general or trivial conversations present throughout the call that affects the quality of information extracted. This fact motivated the development of Conversational Casual Talk Identification system. This system considers causal talk identification as a sentence classification problem using a supervised ensemble model trained on nearly 10,000 sentences. We do not perform any manual tagging – instead we just use the observation that the initial sentences in the call are casual talk and randomly pick equal number of sentences from the rest of the call to be tagged as business talk. Thus, we automatically generate our training set for the Casual Talk Identification system. Manual inspection of a few samples indicated that this heuristic works very well in generating a good training set. The quantitative features derived from each sentence are position of sentence in the transcript, number of tokens, number of stopwords, number of entities, number of person names, and number of geographical location present in a sentence. The system was tuned for high Precision at the cost of some Recall. This means that having some noisy sentences left for the downstream system is acceptable, but the system should not remove any valuable sentences that might contain a key concept or an intent. We achieved 100% Precision and 87% Recall, thus ensuring that none of the downstream systems has to deal with noisy sentences.

### 3.2    Key concept identification

Our system design for extracting key topics discussed during a conversation follows a funnel approach in which we start with almost all possible phrases within a transcript ensuring high Recall and as we move down the funnel, optimize for Precision at the cost of Recall by removing phrases that are not key concepts. The entire process is broken down into 4 modules in which each module focuses on its principal objective by executing various steps to fulfil it. Below are the details for each of the modules.

**Phrase Extraction**

We begin our pipeline following common pre-processing steps of lowercasing the text and removing any noise, present due to source channel. The main objective of this step is to extract phrases that are defined as n-grams ranging from 1 to 5. In the process of phrase extraction, we use certain rules which takes care of the sentence boundary, punctuations and use of any non-alphabetic character in the phrase. This step is the start of the funnel providing us all the potential candidates for the key concepts discussed in the conversation.

**Noise Removal**

This module begins with set of phrases from the phrase extraction module to drop phrases that are highly unlikely to be key topics referred as noise. It involves steps such as (1) checking if the phrase has leading and ending token as stop word (2) tokens of the phrase are from English vocabulary (3) phrase is not part of the casual conversation (refer section 3.1) (4) phrase not part of conversational stop-words to deal with phrases like "gon na" which are commonly used in a conversation. We developed a module that will allow us to automatically identify domain specific conversational stop-words such as "gon na", "like" etc.

In addition to the above rules, Named Entity Recognition (NER) tags of "PERSON"," LOCATION"," QUANTITY", "TIME" types are extracted from the corpus and phrases having high semantic similarity to such tags are dropped as they are most likely to be treated as noise for the present problem. The last step in this module involves dropping phrases that have very low phrase IDF values (which are learnt from entire conversational corpus) and also the average of the token level IDF values are considered for cases where few tokens are very common within a phrase. This module helps to reduce the number of extracted phrases by 50-60%.

**Phrase Normalization**

Most of the phrases post the noise removal module are clean, yet they consist of many repetitions. These repetitions are either due to the variation of the root form of the tokens or have phrases that are very similar to each other. To deal with the 1st form of repetition, phrase lemmatization provides a very effective technique whereas to deal with the 2nd form we use phrase embedding to capture the semantic similarity. There can be many methods to learn the phrase embedding but we relied on the embedding from the BERT [18] language model which was fine-tuned on STSB-NLI dataset for our baseline runs. We also trained the BERT language models on the domain specific vocabulary and our experiments indicated that using the domain trained embeddings had better accuracy in performing phrase normalizations.

The order of operations of these steps are important to remove the redundancy hence we begin by clubbing the phrases which have same root form and this step on an average reduced the phrase set size by 5-10%. Now to group the phrases based on their meaning the remaining set of phrases are ranked based on their frequency and the type of n-gram. Bigrams and trigrams are given higher preference as they provide a bit more information compared to unigrams and are considered ideal candidates for group heads. By the end of this step all the phrases (group heads) are clean and unique and have reduced the phrases set size by another 5-10%.

**Phrase ranking**
This is the last and the final stage of the pipeline in which the ranking of the remaining phrases is carried out. Frequency of phrases combined with the frequency of other similar phrases extracted in the previous module is one of the most important signals considered in the ranking process. Other signal incorporates the (1) POS pattern of the phrase such as multiple occurrences of Noun is very good fit for the abstract topics, (2) location of the phrase in the conversation, (3) number of similar phrases associated with the phrase. These signals help to identify the reason behind the rank for each of the phrase making the solution explainable. Other signals may be added based on any additional data present and corresponding weights need to be altered before the final ranking of the phrases. Phrases at the top are most likely to be the key concepts discussed in the call transcript.

### 3.3 Comprehensive Open Intent Segments Extraction

This paper proposes a novel approach to identify the intent segments in asynchronous text conversations. Generally, the intent is defined as a sequence of words that conveys a specific objective [19,20]. However, we define intent as a group of 2-3 contiguous sentences that can solely convey an idea along with its necessary context. To illustrate, the text "I would like to make a hotel room reservation. And also, if available, make the reservation for a deluxe room" contains only one intent i.e., to reserve a room, possibly a deluxe room. Now, while making the reservation, the person asks "Can you please also tell me the loyalty points I have in my account?". So, this will be a secondary intent that the person asks i.e., information on loyalty points.

The objective of this module is to identify multiple intent segments from text utterances such that each segment can solely convey one whole idea with the necessary context. These can be underlying goals, activities or tasks that a user wants to perform or have performed. The framework is to extract the potential intent sentences, combine the contiguous extracted sentences to form intent segments and then rank the obtained segments. We formulate the extraction of potential intent sentence in two parts: (a) extraction of questions that includes inquiry, probing, leading and open questions (b) extraction of intent conveying sentences.

Question-based Intent Identification: Question Identification module is developed using a rule based unsupervised approach. This approach leverages 5W-1H words and four generalized POS tag and Dependency parser patterns to detect the starting point of the interrogative part, if available, in a sentence. Like most rule-based systems, this approach provided high Precision values but moderate Recall value. To achieve higher Recall value, we developed a hybrid system by incorporating a supervised classifier to the existing system. The classifier was trained on balanced set of 52,000 sentences. Each sentence is represented by a 300-dimensional vector obtained by taking the average of word2vec of each token in a sentence. Other features such as BOW representation of POS tags of each sentence is calculated.

Constraints-based Intent Sentence Identification: Since it is challenging to construct a comprehensive list of all intent indicators [10], we came up with a generalizable list of patterns to identify intent. These include: (i) skip gram matching of the sentence dependency parser with pattern "nsubj aux root" (ii) presence of subjective or possessive adjective followed by word "to" (iii) connected contiguous sentences followed by the above extracted sentences identified by the presence of conjunctions in the beginning of the sentences.

Intent segments are then further formed by combining the contiguous sentences. The ranks of intent segments are further boosted according to number of concepts, questions, and BERT + K-means summarization [21] sentences present in the intent segments. Since the number of intents present in a transcript is not fixed, the system uses a differential cut-off to identify the number of top intents dynamically to provide as output.

### 3.4 Multilabel Document Tagging

Since we are proposing an unsupervised approach to tag transcripts to multiple categories, we leverage the additional data which provides an elaborate explanation for each of the categories in formal English text. We begin our effort to tag the transcripts to multiple categories by representing categories into vector forms, by using the associated supporting docs. Below are the steps followed for vectorization of text:

Vocabulary creation: For each category we combine all the supporting document's text and follow general pre-processing steps for cleaning the text such as lowercasing the text, stop-word removal, lemmatization etc. We also incorporate some of the business knowledge in selecting sections of these documents which are category specific compared to general sections, to use tokens that are relevant to the category.

Feature Selection: In this step we identify the most discriminatory tokens which are selected from the set of tokens that were created at a category level in the previous step. We are using chi-sq methods and mutual information to select the features that will help in removing the noisy feature and reduce the feature set without information loss.

Vector representation: Once we had the final features, we experimented with different methods to represent each category using Bag of Words, Binary, TF-IDF vectorization. TF-IDF vectors for each category gave us the best results based on the evaluation metric for the process (refer to section 4.3).

Once we have represented the categories in vectorized form using the above steps, we transformed the transcripts to the same dimensional space as of categories. We used cosine similarity to identify the nearest categories for each of the transcript document.

## 4 Result

To evaluate the effectiveness of our 1st two systems we used qualitative metrics such as Valuable Concept Percentage (VCP) & Valuable Intent Percentage (VIP) and quantitative metrics such as Precision, Recall and F1 score against a sample of ground truth manually tagged data referred as golden set. We introduced qualitative metric for evaluation as there will be certain concepts and intent which the domain expert would have missed while tagging and this metric effectively captures that aspect. For the evaluation of our last method, we used dynamic Precision for our top 2 predicted results, details of which are mentioned in the below section. We used tagged data only to measure the effectiveness of our methods as all of our proposed solution are unsupervised in nature.

### 4.1 Key concept extraction metrics

We considered fixed number of output (top 10) from the pipeline as the predicted output for each of the transcript and matched with the available ground truth (user tagged data). As the problem here is fundamentally different from the usual classification problem, we changed the methodology for calculation of Precision and Recall keeping the formula same. The steps involved are (1) converting the ground truth and predicted result to their base form (2) removing any general English stop-word present in ground truth and predicted result (3) considering partial match (token level) instead of exact match.

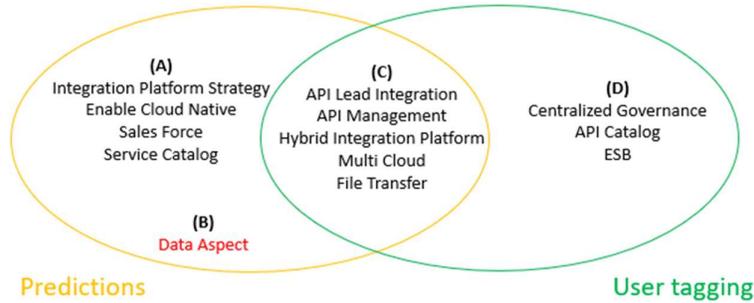

$$Precision = \frac{\#\ partial\ match}{\#\ predicted\ output} = \frac{C}{(A + B + C)}$$

$$Recall = \frac{\#\ partial\ match}{\#ground\ truth\ provided} = \frac{C}{(C + D)}$$

**Fig. 1: Example demonstrating key concept extraction metrics calculation**

For calculation of VCP, the predicted result for each of the transcripts are tagged by the domain expert into 2 categories – useful & noisy and the proposed method achieved 85% concepts in the useful bucket. Since the number of ground truth for each tagged transcript varied and was usually below 10, Precision score was ~ 40% i.e. lower than the Recall which was ~ 55%. Based on the discussed results the solution on an average had (1) 5 concepts that matched with the tagged ground truth (2) 1-2 noisy concepts (3) extracted 3-4 valuable concepts which the domain expert missed, for a given top 10 result.

### 4.2 Comprehensive Open Intent Segments Extraction metrics

Similar to the above discussed VCP metric, we formulated VIP metric to measure the valuable number of intent segments. The predicted results for each transcript are tagged by the domain expert – useful and noisy and the proposed method achieved 82% of useful intent segments. Since a greater number of sentences will have high probability of having a useful intent, we also studied the number of sentences in the output intent segments. We observed that 75% of the intents have less than or equal to 3 sentences and the maximum number of sentences in any intent being 6. This shows that the intent segments are concise and consider only the essential contextual sentences in each intent segment. We have considered each sentence as a single entity for calculation of the F1 score.

$$Precison = \frac{\text{\# exact sentence match}}{\text{\# predicted sentences output}} = \frac{C}{(A + B + C)}$$

$$Recall = \frac{\text{\# exact sentence match}}{\text{\#ground truth sentences provided}} = \frac{C}{(C + D)}$$

Using the above method, we achieved 45% Recall and 25% Precision. The above results can be summarized as (a) an average of 3 intents are predicted for each transcript and 3-4 sentences predicted per intent segment (c) 1 intent segment matched with the user tagging (d) 1 intent segment is valuable but not tagged by the user (e) within the segment, 1-2 noisy sentences across 3 predicted intent segments.

### 4.3 Multi-document Tagging metrics

For multi-document tagging, we are using the Recall at k which is defined as the ratio of output category tags that are relevant to the document to the relevant output tags. Although, there can be multiple category tags to each document, the ground truth contained only one category corresponding to one document that was tagged by domain expert. Using the above approach, we achieved 85% Recall@2 value.

$$Recall@k = \frac{\text{\# of output category tags @k that are relevant}}{\text{\# of relevant output category tags @k}}$$

## 5    Conclusion

This paper presents an unsupervised approach to extract key topics and intents discussed during a call and classify the transcript into predefined categories using unsupervised method. The key concept extraction algorithm is customizable, scalable and explains the major contributing factors for each extracted topic. To provide more information about the call we presented open intent recognition method to identify the intent of the call and dynamically output multiple intents for each transcript. Both information extraction methods leverage the BERT language model trained on the domain data that provides improved results over pre-trained models. To tag each call transcript data into different categories we discussed the multi label tagging approach which works on untagged data at scale. We further discussed the need to develop new metrics to understand the true performance of the developed systems. All our proposed methods don't require any training data and achieves near human performance.